# Heavy holes – precursor to superconductivity in antiferromagnetic CeIn$_3$

Suchitra E. Sebastian[1], N. Harrison[2], C. D. Batista[2], S. A. Trugman[2], V. Fanelli[2], M. Jaime[2], T. P. Murphy[3], E. C. Palm[3], H. Harima[4] and T. Ebihara[5*]

Numerous phenomenological parallels have been drawn between *f*- and *d*-electron systems in an attempt to understand their display of unconventional superconductivity[1,2,3,4,5,6,7,8]. The microscopics of how electrons evolve from participation in large moment antiferromagnetism to superconductivity in these systems[9], however, remains a mystery. Knowing the origin of Cooper paired electrons in momentum space is a crucial prerequisite for understanding the pairing mechanism. Of especial interest are pressure-induced superconductors CeIn$_3$ and CeRhIn$_5$[10,11] in which disparate magnetic and superconducting orders apparently coexist – arising from within the same *f*-electron degrees of freedom. Here we present ambient pressure quantum oscillation measurements on CeIn$_3$ that crucially identify the electronic structure[12] – potentially similar to high temperature superconductors[13,14,15,16,17]. Heavy pockets of *f*-character are revealed in CeIn$_3$, undergoing an unexpected effective mass divergence well before the antiferromagnetic critical field. We thus uncover the softening of a branch of quasiparticle excitations located away from the traditional spin-fluctuation dominated antiferromagnetic quantum critical point. The observed Fermi surface of dispersive *f*-electrons in CeIn$_3$ could potentially explain the emergence of Cooper pairs from within a strong moment antiferromagnet.

[*] [1]Cavendish Laboratory, University of Cambridge, Madingley Road, Cambridge CB3 0HE, UK

[2]National High Magnetic Field Laboratory, Los Alamos National Laboratory, MS E536, Los Alamos, New Mexico 87545

[3]National High Magnetic Field Laboratory, 1800 E. Paul Dirac Drive, Tallahassee, Florida 32310

[4]Department of Physics, Kobe University, Kobe 657-8501, Japan

[5]Department of Physics, Shizuoka University, Shizuoka 422-8529, Japan

With a Neel ordering temperature of $T_N \approx 10$ K [Ref.8], CeIn$_3$ belongs to a family of Ce-based antiferromagnets that become superconducting under applied pressure. At first glance, CeIn$_3$ at ambient pressure appears to exhibit all the hallmarks of a localised antiferromagnet. As our magnetic oscillation measurements (Fig. 1) confirm, the Fermi surface of CeIn$_3$ is very similar to that of LuIn$_3$ and LaIn$_3$ [Ref. 18] (in which the *f*-electron shells are filled and empty respectively); the measured d- and k-orbits closely correspond in magnetic field-orientation dependence and volume to identical sheets measured (and calculated) for these nonmagnetic compounds [Ref. 12,13,14,15]. Further, the weak mass enhancement of the observed orbits[†], and large staggered moment between 70 and 90 % of the local Ce moment value (a $\Gamma_7$ doublet in CeIn$_3$)[19,20] point to weak hybridisation of the conduction with the *f*-electrons, which minimally participate in the Fermi surface. We are then left with a question as to the electronic origin of superconductivity under pressure in CeIn$_3$ and not its nonmagnetic analogues.

Indications of the unconventional nature of *f*-electron participation in the Fermi surface of CeIn$_3$ appear on cooling to dilution refrigerator temperatures T << 500 mK. The light conduction electron Fermi surface remains largely unchanged (Supporting Information). However, we find it to be accompanied by new r-orbits with unexpectedly large effective masses $m^* \sim 20 m_e$ in magnetic fields between 10 and 20T, which are absent in the non-magnetic analogues. The topology of the corresponding sections of Fermi surface is deduced from the magnetic field-orientation dependent frequencies, a few branches of which were previously observed[20]. Our measurements reveal a classic three branch form corresponding to spatially separated ellipsoids of revolution situated along the <111> directions of a cubic system[12]. The new section of Fermi surface is thus associated with 8 oblate ellipsoids at **k** = <*k,k,k*> where $k = (0.5 \pm 0.1)\pi$ (shown in the lower inset to Fig. 1a), which dominate the thermodynamic mass. In contrast to the conventional picture for the largely uniform Fermi surface mass enhancement by *f-d* hybridisation, we argue that selective mass enhancement of ellipsoidal pockets disconnected from the conduction electron Fermi surface (Fig. 1c) points to component quasiparticles chiefly originating within the *f*-band. These heavy ellipsoidal pockets are chiefly responsible for the previously unexplained large residual Sommerfeld coefficient[12] reported within the antiferromagnetic phase of CeIn$_3$, $\gamma \approx 120$ mJmol$^{-1}$K$^{-2}$ [Ref. 27] - greatly exceeding that of nonmagnetic LuIn$_3$ and LaIn$_3$ (5.3 mJmol$^{-1}$K$^{-2}$).

---

[†] Some mass enhancement is observed in concentrated regions along the <1,1,1> direction[20], which we explain in the Supporting Information.

In order to understand the unique ellipsoidal Fermi surface topology, we begin in the non interacting limit described by the weak coupling picture. Some evidence for a Fermi surface sheet of chiefly itinerant *f*-electrons has been observed in dHvA experiments on the paramagnetic phase of CeIn$_3$ under pressure[19]. A similar Fermi surface topology would be expected from a less than half filled band of *f*-character (Fig. 1a – upper inset) within a tight binding approximation, as seen from Local Density Approximation (LDA) band structure calculations in the paramagnetic phase of CeIn$_3$ [Ref. 21]. The consequence of band folding of such a surface in the antiferromagnetic regime would be to create ellipsoidal hole pockets at $\mathbf{k} = (\pi/2, \pi/2, \pi/2)$ locations in momentum space (figs 2c, d), similar to those observed in CeIn$_3$. The two-dimensional projection shown in Fig. 2 further reveals the topology of the three dimensional pockets in CeIn$_3$ to bear a close resemblance to the recently observed hole pockets in the two-dimensional Fermi surface of high temperature superconductors. Within the weak coupling picture, weak Bragg scattering at the antiferromagnetic zone boundaries would imply an occupation number in momentum space $\left\langle n_{fk} \right\rangle = \left\langle \sum_{\sigma} f_{k\sigma}^{+} f_{k\sigma} \right\rangle \lesssim 2$ within the original Fermi surface, and $\left\langle n_{fk} \right\rangle \gtrsim 0$ outside.

A similar Fermi surface topology is obtained for the more physical large Coulomb repulsion limit applicable to the large moment antiferromagnet CeIn$_3$, by treatment within a strong coupling picture. In this representation, the less than half filled band of chiefly $f$-character is modelled as a background of antiferromagnetically aligned local moments interspersed with a finite concentration of holes. The interplay between the kinetic energy ($t$) and electron-electron interaction (approximated by the on-site Coulomb interaction $U$) of this band can be treated within a single-band Hubbard model. A similar ellipsoidal Fermi surface topology is obtained, while the large Coulomb repulsion means that double occupancy is inhibited and strong scattering by the antiferromagnetic periodic potential results in an occupation number $\left\langle n_{fk} \right\rangle \sim 1$ throughout the Brillouin zone. Holes aggregate about the ($\pm\pi/2, \pm\pi/2, \pm\pi/2$) points in momentum space (Fig. 2e), which are nested by the antiferromagnetic ordering wavevector Q = ($\pi,\pi,\pi$) at half filling[20,23]. The interior of the resulting pockets has a lower occupation number, approaching zero within the mean field approximation. Within this picture, 'heaviness' results predominantly from the magnetic distortion accompanying hole motion in an antiferromagnetic 'background'[24].

The heavy ellipsoidal pockets in CeIn$_3$ show further surprises in their magnetic field tuned properties. Whilst a modification in Fermi surface properties would be natural in the vicinity of the critical field $B_c \sim 61$T where antiferromagnetism is suppressed, we find that astonishingly, the ellipsoidal mass diverges at an intermediate field $B_f \approx 0.7 B_c \approx 40$ T accompanied by an upturn in frequency. Divergences in the effective masses of the r1$_{110}$ and r2$_{110}$ orbits are observed as the field approaches 40 T (Fig. 2a), well inside the antiferromagnetic phase and unaccompanied by features in the spin susceptibility[24]. We show that these signatures can be interpreted in terms of magnetic field tuned correlation effects within a single band of chiefly $f$-character.

In the non-interacting picture, a magnetic field applied perpendicular to the Néel axis would result in the two overlapping bands moving in opposite directions toward the left and right, causing the ellipsoids to migrate toward the zone centre (upper right quadrant of the original Brillouin zone shown in Figs. 2d and f). Equivalently, in the single band Hubbard model, the migration of ellipsoids toward the zone centre with magnetic field (Fig. 3) is associated with the transformation of background periodic potential from antiferromagnetic to ferromagnetic. In the ferromagnetic (saturated) limit, the concentration of holes moves in a background of all parallel aligned spins, from which we readily see that the quasiparticle dispersion minimum moves from $\mathbf{k}=(\pi/2, \pi/2, \pi/2)$ at low fields to $\mathbf{k} = (0,0,0)$ in the high field limit. In essence, the dominant intra-sublattice hopping process in the antiferromagnetic regime evolves to a dominant nearest neighbour hopping process in the ferromagnetic regime, resulting in a migration of the quasiparticle dispersion minimum.

The strongly-coupled single $f$-band picture we propose can reproduce the magnetic field-induced features in mass and frequency of the ellipsoidal pockets near $B_f$ in CeIn$_3$. We discuss two ways in which the evolution of a single band with magnetic field can contribute to the effective mass divergence of the ellipsoidal pockets in the vicinity of $B_f$; and investigate the likely associated softening of quasiparticle excitations in this region. Firstly, enroute their migration toward zone centre, the eight ellipsoids coalesce at an intermediate field in a topological transformation known as the 'Lifshitz Transition'[25], where the quasiparticle band flattens (Fig. 3). A change in sign of the quadratic dispersion term occurs at this intermediate field due to phase cancellation between two second order hopping processes that differ by the exchange of two fermions (Supporting Information, Supporting Fig.5). The consequent vanishing of the quadratic dispersion term contributes to the effective mass divergence, and the field at which this occurs would coincide with $B_f \approx 40$T for values of parameters $U/t = 8$, $t'' = -0.022|t|$ and $t = 70$ meV in the $t$-$t'$-$t''$-$J$ representation of the single band Hubbard model including next nearest neighbour coupling (shown in Fig. 4a). The second consequence of the magnetic field is to lower the single band dispersion maximum, resulting in the shrinking of the $f$-hole pockets and their ultimate depopulation (Fig. 3). The experimentally observed sharp upturn in frequencies is associated with an increasing rate of Landau level depopulation $F = -B^2 \partial \nu / \partial B$ [See Supporting Information, Ref. 26] with $B$ (where $\nu = hA_k/4\pi^2 eB$ is the Landau level filling factor). The associated volume dependence of the ellipsoidal pockets on magnetic field is shown in the inset to Fig. 4b, identifying the depopulation field where the pockets completely vanish as $B_d \approx 41.3$ T, in the close vicinity of $B_f$. The loss of $f$-electrons from the Fermi surface at $B_d$ also means that the charge associated with these particles is no longer bound within the Fermi surface. The softening of the charge mode at the depopulation field may therefore be associated with the divergence in $f$-hole pocket mass well before the critical field $B_c$ where the spin mode softens.

The contribution of the heavy $f$-hole pockets at $\mathbf{k} = (\pi/2,\pi/2,\pi/2)$ to the electronic heat capacity $\gamma T$ within the antiferromagnetic phase dwarfs that of the conduction bands. An integral $\gamma = k_B^2/6h \int dS/|v_F|$ over the Fermi surface of the $f$-hole pockets (where $\mathbf{v}_F = 2\pi \nabla_k \varepsilon / h$ is the Fermi velocity) reveals that they alone account for at least 80 % of the experimental electronic heat capacity at zero field[27], revealing a concentration of density-of-states at the hole pockets in the antiferromagnetic phase. With the emergence of superconductivity from antiferromagnetism under pressure[9,11], the Fermi surface of the regular conduction bands remains essentially unchanged[19], indicating the likely formation of Cooper pairs almost exclusively at or near these hole pockets situated at $\mathbf{k} = (\pi/2,\pi/2,\pi/2)$.

The underdoped state of high temperature superconductors exhibits small pockets at the equivalent $\mathbf{k} = (\pi/2,\pi/2, q)$ locations, as revealed by quantum oscillations[13,14,15,16,17] and photoemission[29]. Whereas previous studies have suggested phenomenological commonalities between $d$- and $f$- superconductors[4,10], here we show a potential parallelism at the microscopic level between the electronic structures of the precursor states to superconductivity[5,28,29]. Doped Mott insulator physics is mimicked in $CeIn_3$, resulting in a three-dimensional Fermi surface with topological similarities to the two-dimensional cuprate Fermi surface. The effective charge reservoir provided by the conduction bands in $CeIn_3$ could play a similar role to the magnetically inert Oxygen chains in the YBCO cuprates. The emergence of a new magnetic field tuned mass divergence of the hole pockets in $CeIn_3$ points to a softening of a distinct class of quasiparticles associated with these pockets of $f$-character. The separation of critical fields in $CeIn_3$ where the two distinct modes (spin and potentially charge) soften appears to contrast with recent experimental findings in $CeRhIn_5$ under pressure, where the two are reported to coincide [Ref. 30]. Further experiments on the hole pockets in $CeIn_3$ by tuning along the pressure axis will be of interest in exploring where the softening of the new mode occurs in relation to superconductivity and the suppression of antiferromagnetism. $CeIn_3$ provides us with a structurally simple model system to explore the relationship between the evolution of the hole pocket regions in momentum space under pressure (or doping), and the concomitant appearance of superconductivity.

Acknowledgements

This work was performed under the auspices of the National Science Foundation, the Department of Energy (US) and Florida State. Support was provided by Grant-in-Aid for Scientific Research on priority Areas, 'High Field Spin Science in 100T' and MEXT. We thank G. G. Lonzarich and P.B.Littlewood for valuable inputs, and R. D. McDonald and A. C. H. Cheung for technical assistance with simulations. SES acknowledges support from the Mustard Seed Foundation, Institute for Complex Adaptive Matter, and Trinity College (Cambridge University).

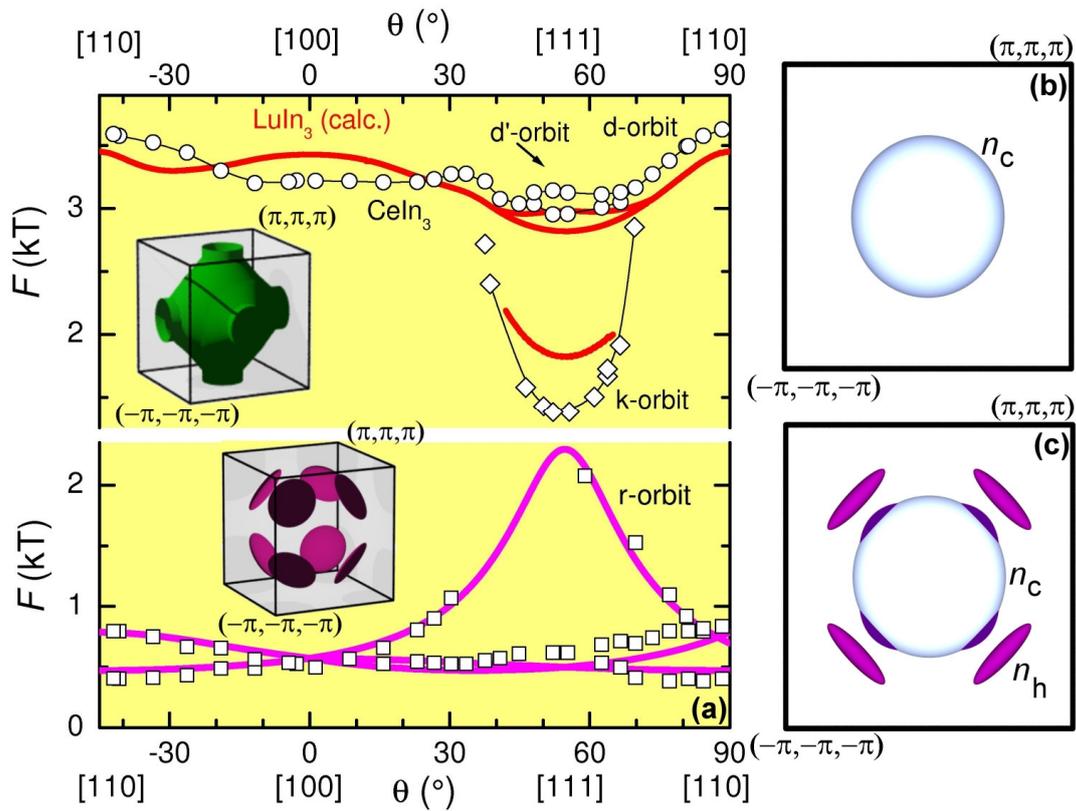

Fig. 1: Fermi surfaces of CeIn$_3$. **a.** measured d, k- and r- orbit quantum oscillation frequencies $F$ (obtained by magnetic torque), related to the Fermi surface (FS) cross-sectional area $A_k$ [Supporting Information], versus the orientation of **H**. Electrons undergoing cyclotron motion in real-space circumnavigate the FS in momentum space **k**. Measured d- and k- band frequencies are similar to those in nonmagnetic LuIn$_3$ (red lines) calculated by an augmented plane wave method, whereas r-orbit frequencies, found only in CeIn$_3$, correspond to simulations (purple lines) of the 8 ellipsoidal pockets depicted in the lower inset. The upper inset shows a less than half filled $f$-electron Fermi surface sheet from LDA calculations in the paramagnetic phase of CeIn$_3$ [Ref. 21]. (b) Schematic representation of the conduction electron Fermi surface (d-sheet) expected within the strong moment Kondo lattice model. In this picture, any mass enhancement would occur due to conduction-$f$ electron hybridisation over the entire conduction electron Fermi surface. (c) Schematic for unconventional participation of $f$-electrons in Fermi surface indicated in CeIn$_3$. A distinct heavy surface (argued to be of principally $f$-

electron character) appears, causing an enhancement of the conduction electron density of states only in the regions of conduction-$f$ electron band proximity.

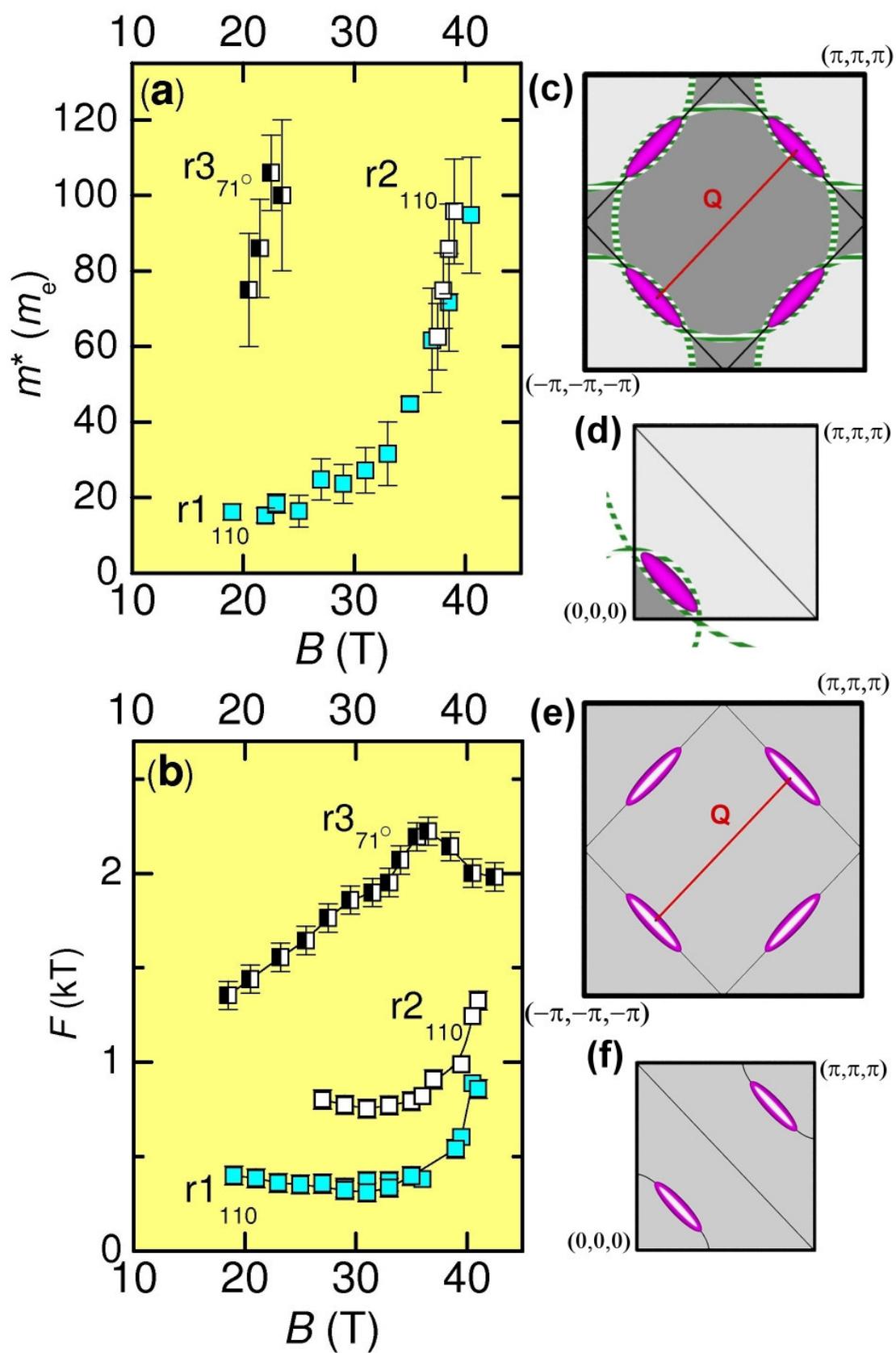

Fig. 2: Field-dependence of the ellipsoidal r-orbits observed in CeIn$_3$, evidenced by torque measurements made using a portable dilution refrigerator in the 45 tesla hybrid magnet at NHMFL, Tallahassee [Supporting Information]. **a**. Effective masses obtained on fitting the temperature dependent quantum oscillation amplitude to the Lifshitz-Kosevich theoretical form Amp($T$) = Amp$_0$ $X$ / sinh $X$ where   $X = 4\pi^3 k_B m^* T / ehB$ [Ref. 31] for 33 < $T$ < 500 mK (~5% uncertainty in $T$). A Hanning window is used in Fourier transforming over 5-8 T intervals. Symbols denote frequencies plotted in **b**. Subscripts denote the orientation of **H** in Fig. 1. **c**. Weak coupling picture: 2d projection of $f$-electron surface of topology shown in **Fig. 1** (upper inset) folded at the antiferromagnetic zone boundary to give rise to hole pockets. Within this picture, an $f$-electron occupancy $\left\langle n_{fk} \right\rangle \lesssim 2$ inside the original Fermi surface and $\left\langle n_{fk} \right\rangle \gtrsim 0$ outside (indicated by the shading) would be implied due to weak Bragg scattering. **d**. Migration of ellipsoids toward the zone centre due to Zeeman splitting in the weak coupling picture **e**. Strong coupling picture: strong scattering leads to an $f$-electron occupancy of $\left\langle n_{fk} \right\rangle$ ~ 1 everywhere in the Brillouin zone. Away from half filling, the residual holes aggregate at the nesting wave-vector ends ($\pm\pi/2$, $\pm\pi/2$, $\pm\pi/2$). The occupancy number in the interior of the resulting pockets drops to a lower value approaching zero in the mean field limit. **f**. Migration of ellipsoidal pockets due to motion of bands in opposing directions with field in the strong coupling picture.

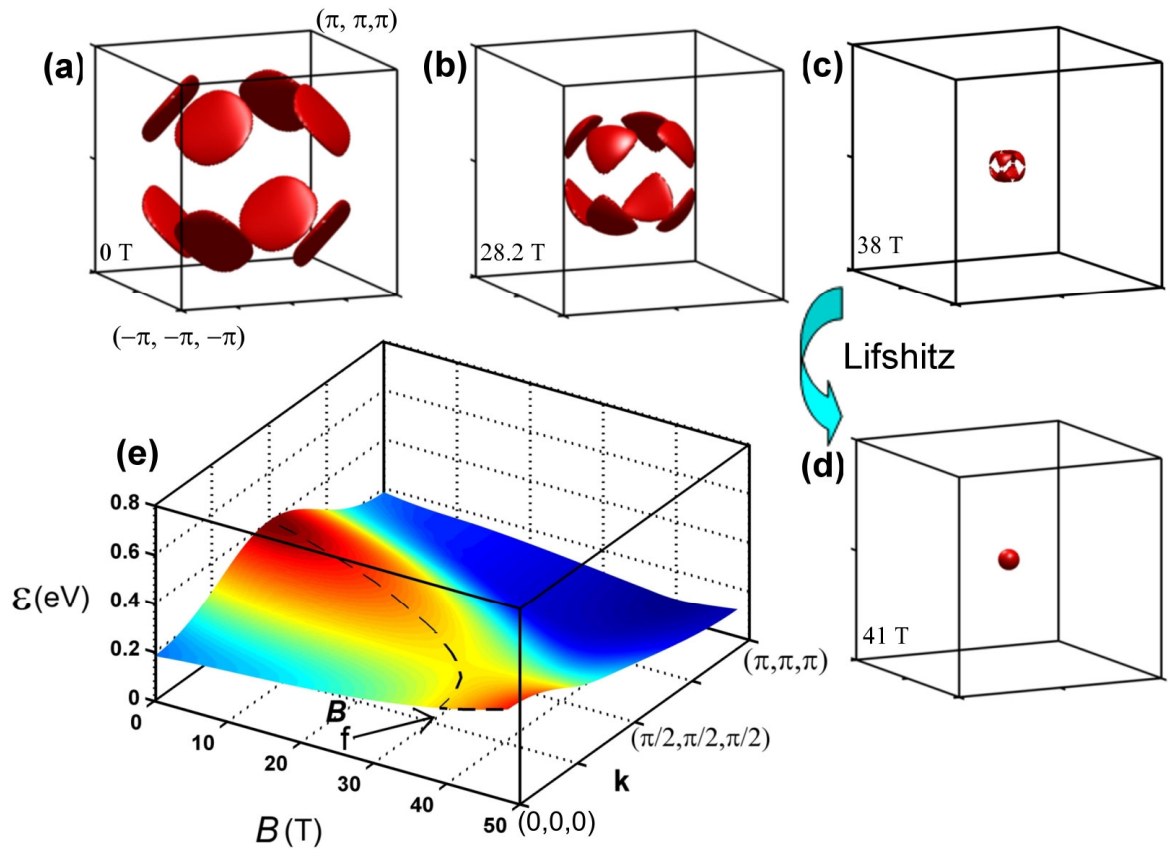

Fig. 3: Schematic Fermi surface and *f*-electron dispersion of CeIn$_3$. **(a-d)** Theoretical *f*-hole Fermi surface topologies at different magnetic field simulated using a strong coupling Hubbard model with parameters $U/t$ = 8, *t'* = -0.022|*t*| and $t$ = 70 meV – a Lifshitz transition is indicated in the vicinity of 40T. **e.** Calculation of the dispersion along <111> using the strong-coupling expansion of the Hubbard model. The dispersion maximum (hence pocket location) migrates from ($\pi/2,\pi/2,\pi/2$) to (0,0,0), following the dotted line, as the antiferromagnetic background is polarised by $B$. Band flattening near the dispersion maximum at $B_f$ ~ 40 T is responsible for the divergent effective mass and the Lifshitz transition (see Fig. 4a).

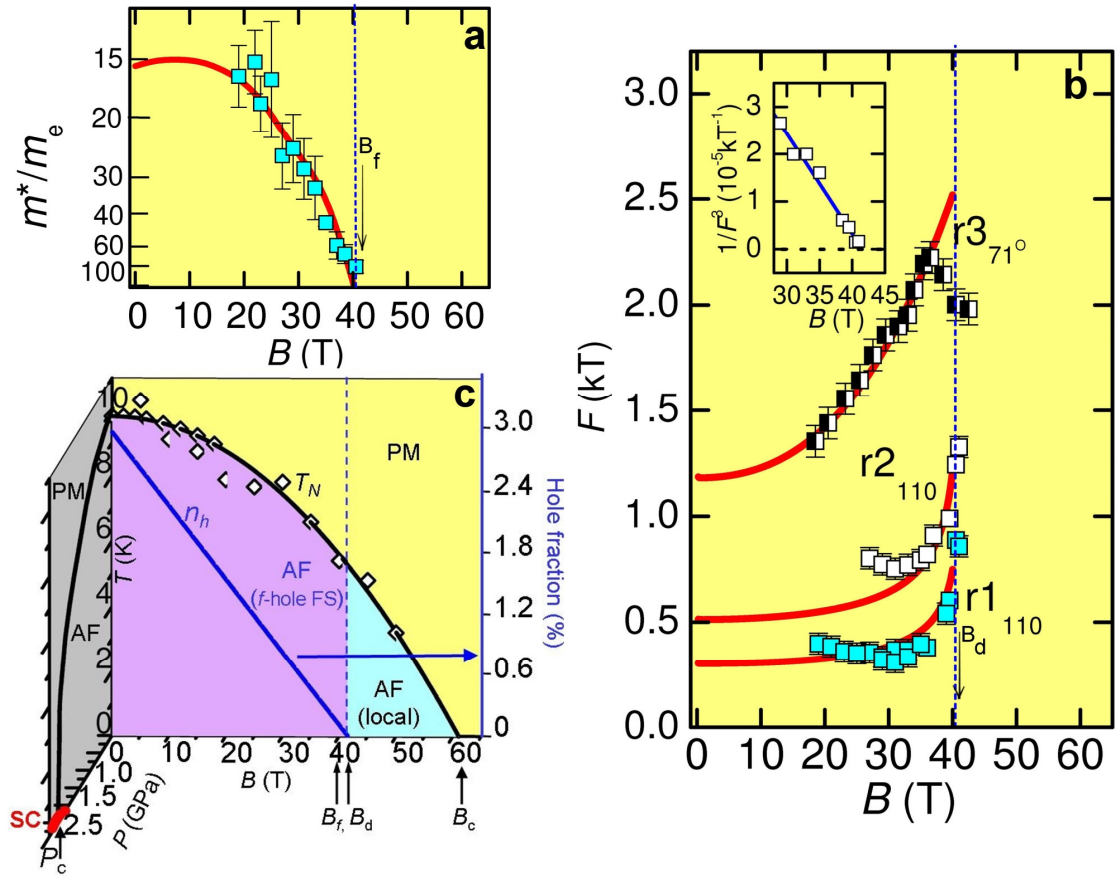

Fig. 4: Intermediate transition field from experiment and theory. **a** $1/m^*$ for the r1$_{110}$ frequency (cyan squares), in agreement with theoretical prediction (red lines), vanishes at $B_f \approx 0.7B_c$ - suggesting $m^*$ divergence within the antiferromagnetic phase[24] **b.** r-frequencies in Fig. 2b (squares) compared with theoretical $F = - B^2\partial(hA_k/4\pi^2eB)/\partial B$ [See Supporting Information, Ref. 26] simulations (red lines) for an $f$-hole Fermi volume of $V = 0.375 \times (1-B/B_d)\%$ per ellipsoid (Brillouin zone percentage). Linear depopulation is evidenced in the inset by the linear dependence of $1/F^3$ on $B$ [Supporting Information]. The intercept yields the depopulation field $B_d \approx 41.3$ T (where $v = 0$). The opposing $B$-dependent trends of $F$ and the **k**-space area $A_k$ constitute an extreme case of "back reflection"[26]. **c** Phase diagram to indicate the intermediate field within the antiferromagnetic phase boundary (blue line and diamond symbols) at which mass divergence and ellipsoid depopulation accompany a transition from a regime with an $f$-hole Fermi surface to a local antiferromagnetic regime.

**Supporting Information**

**Fermi-surface topology**

 Fermi surface topology corresponding to the measured d- and k- band frequencies in CeIn$_3$ as calculated by the augmented plane wave method for nonmagnetic LuIn$_3$ is shown in Supporting Fig. 1a,b (associated frequencies shown by red lines in Fig. 1a in the main text). Fig. 1c,d show schematics of the r-orbit unique to CeIn$_3$ for two fixed orientations of H (aligned ⊥ to the page).

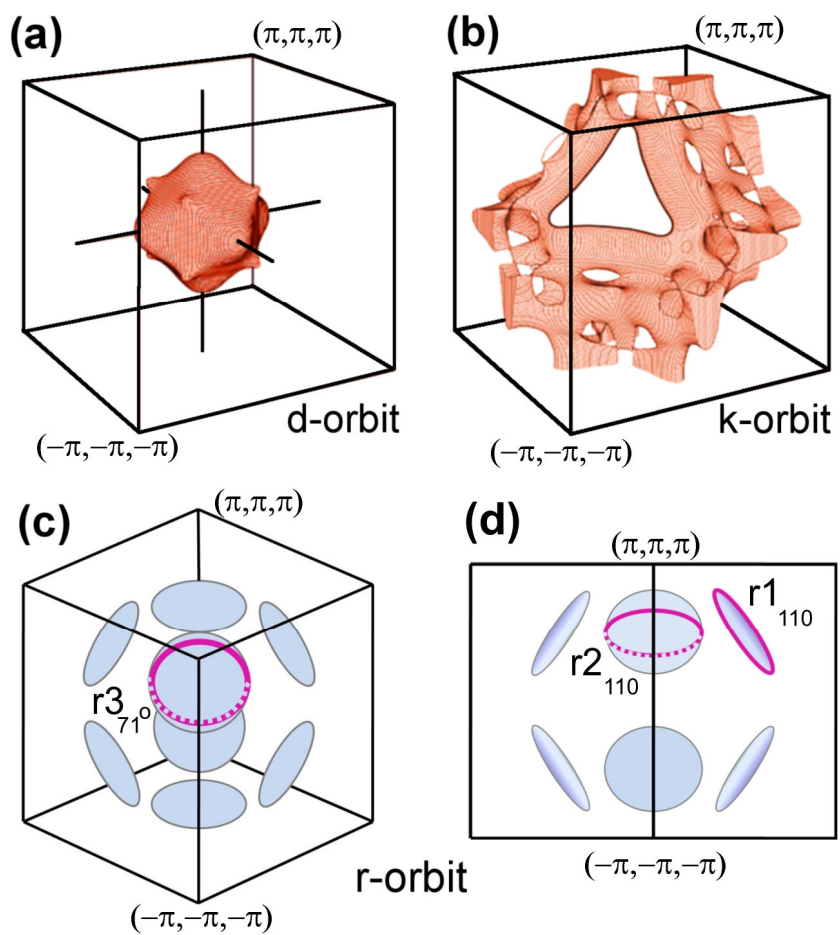

Supporting Figure 1: (a,b) Topology of the d- and k- Fermi surface sheets in CeIn$_3$, identical to those measured and calculated in nonmagnetic LuIn$_3$. (c,d) Cross-sections of the ellipsoidal r-orbit observed only in CeIn$_3$.

## Field-evolution of conduction electron orbits

The field-induced transfer of holes between sheets (in accordance with Luttinger's theorem) causes the $d_{110}$ and $d'_{111}$ electron orbit frequencies to show a similar (albeit weaker) magnetic field-dependent trend to those of the r-orbits in Fig. 2b – shown in supplementary Figure 2. The reduction in area occurs mostly at the humps (or 'hot spots') on the d-sheet [17,20], also pointing along <111> (shown in Fig. 1b) where their dispersion in **k** is weakest. The enhanced effective masses of the $d_{110}$ and $d'_{111}$ orbits passing over these humps in Fig. 2a can be explained by a weak hybridisation between the *f*-holes and d-sheet electron carriers that becomes increasingly relevant as the *f*-hole dispersion maximum shifts along <111> towards the humps with increasing field.

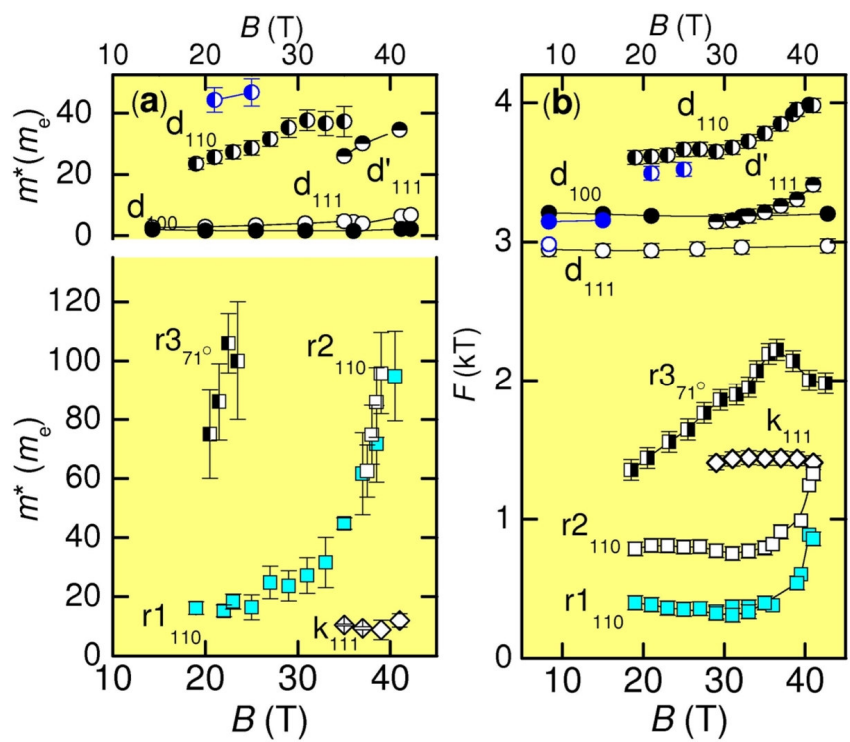

Supporting Figure 2: Field-dependence of the d- and k- Fermi surface sheets of CeIn$_3$ shown alongside that of the r-sheet. Blue symbols may reveal two separate exchange-split spin components of the d-frequency.

**Magnetic quantum oscillation measurement techniques**

Quantum oscillation measurements are performed in CeIn$_3$ in a magnetic field range up to 45 T at temperatures down to 33 mK using the magnetic torque method. The torque cantilever magnetometer is mounted on a platform rotatable about the axis of torque to enable field-orientation dependent measurements. The data shown in Figures 1 & 2 were obtained within ~2$^o$ of symmetry axes. The effective mass associated with each frequency was obtained by fitting the temperature dependent dHvA amplitudes to the Lifshitz Kosevitch (LK) formula[30], assuming Fermi liquid behavior (examples shown in Supporting Figure 3).

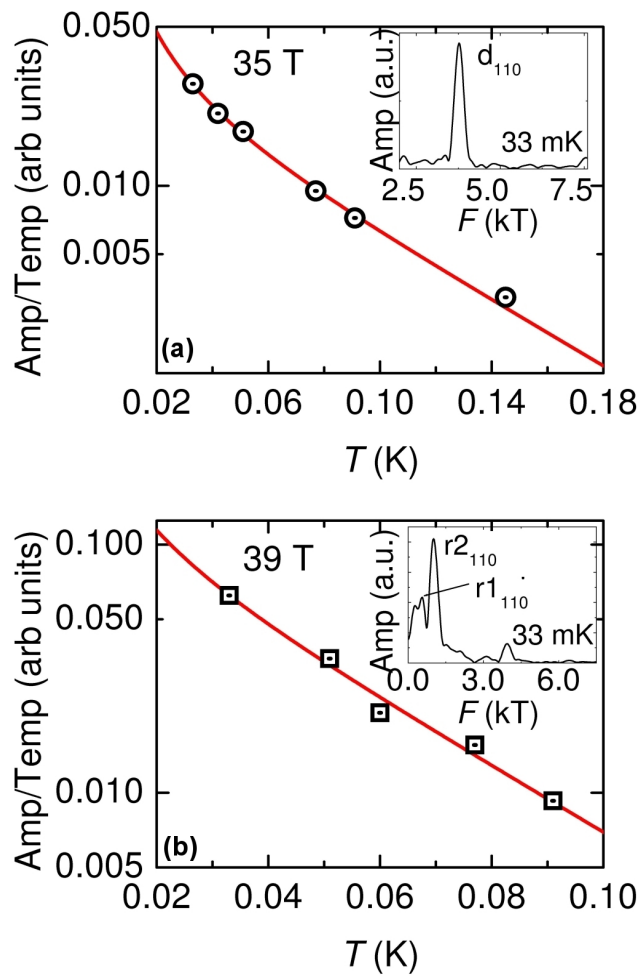

Supporting Fig. 3: Example Lifshitz-Kosevich fits to the amplitude of the dHvA frequency to obtain the effective mass. (a) shows the fit corresponding to the d-orbit for **H** || <110>, yielding $|m^*|/m_e$ = 38 ± 2 in a magnetic field of 35 T. The inset to (a) shows the frequency corresponding to the d-orbit in the dHvA FFT spectrum. (b) shows the fit corresponding to the r2-orbit for **H** || <110>, yielding $|m^*|/|m_e|$ = 86 ± 10 in a magnetic field of 39 T. The inset shows the frequency corresponding to the r-orbits in the dHvA FFT spectrum.

**Analysis method relating the measured frequency to the orbital cross sectional area of the Fermi surface**

$F$ is the rate at which consecutive Landau levels (filling factor ν) depopulate as their degeneracy increases with $B \approx \mu_0|\mathbf{H}|$. Differentially $F = \partial\nu/\partial(1/B)$. Hence, we obtain $F = - B^2\partial(hA_k/4\pi^2eB)/\partial B \equiv h(A_k - B\partial A_k/\partial B)/4\pi^2e$ (2). Effectively, the measured frequency ($F$) represents a projection of the field-dependent area ($A_k$) (i.e. an extrapolation of the tangent to the origin). For a non-interacting paramagnetic Fermi surface in which $A_k$ is linearly dependent on $B$, this reduces to the Onsager equation $F = hA_0/4\pi^2e$ (used at low $B$ in Fig. 1), where $A_0$ is the cross-sectional area at $B = 0$. For our case where $A_k$ is strongly non-linearly dependent on $B$, however, the field-dependence of $F$ in fact translates to a field-dependence of $A_k$ in the opposite direction (shown for the measured data in Supporting Fig. 4). In particular, for the case of a ellipsoidal Fermi surface whose *volume V* decreases linearly with $B$, inserting $A_k = (hA_0/4\pi^2e) (1-B/B_d)^{2/3} \propto V^{2/3}$ into the above expression yields $F = (hA_0/4\pi^2e)(1-B/3B_d)/(1-B/B_d)^{1/3}$; closely approximated by $1/F^3 \propto (1-B/B_d)$ on the approach to the depopulation field $B_d$, providing an experimental means for identifying *f*-hole pocket depopulation in Fig. 4b.

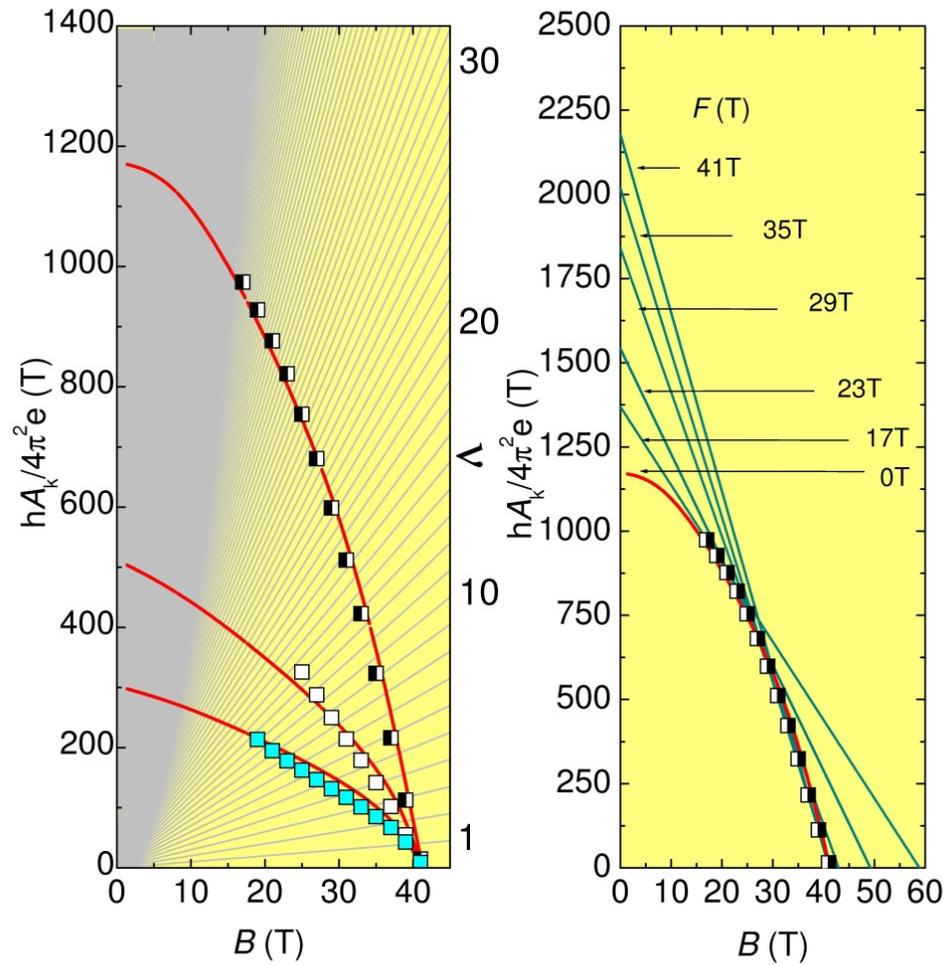

Supporting Fig. 4: (a) Comparison of the field dependence of the Fermi surface cross-sectional areas $A_k$ corresponding to the measured r1$_{110}$, r2$_{110}$ and r3$_{71}^{o}$ frequencies shown in Fig. 2b in the main text (squares) with theoretical simulations (red lines) explained in the main text. The rapid shrinking of the cross-sectional area with magnetic field explains the rise of measured frequency with field – as shown by a back projection[27] represented in (b).

## Strong coupling Hubbard model – details of the model employed

Strong on-site Coulomb interactions between $4f$-orbitals accompanied by their hybridisation with broad conduction and empty bands leads to $f$-itinerancy as well as effective Kondo and $f$-$f$ antiferromagnetic interactions (RKKY and super-exchange). If localised (one $f$-electron per site), low energy properties are entirely described by the $f$-spin degrees of freedom (Kondo Lattice Model)[7,8]. In the local moment antiferromagnetic phase, the $f$-degrees of freedom can be "isolated" and described by an effective Heisenberg Hamiltonian because the Kondo interaction is irrelevant. Transition to a regime where a small fraction of $f$-electrons are promoted to the conduction band (rendering the valence less than integral) is usually accompanied by the formation of a paramagnetic state. While CeIn$_3$ is a prominent exception that remains a large moment antiferromagnet in this regime, Kondo exchange is still relatively unimportant – enabling us to adopt a single band description to describe the $f$-degrees of freedom. In particular, an effective $f$-$f$ hopping process that appears to second order in the hybridisation is the minimal term to describe $f$-hole itinerancy[26].

We show in the main text that the observed field-induced divergence of the effective mass in the antiferromagnetic phase of CeIn$_3$ can be captured by this model. In Supporting Fig. 5, we show the field-induced change in the relative strength of different hoppings responsible for the local flattening of the itinerant $f$-hole band around the dispersion maximum (Fig. 3e in the main text), resulting in an effective mass divergence. Magnetic field induced spin polarisation induces ferromagnetic correlations that destructively interfere with antiferromagnetic correlations to an extent that evolves with $B$.

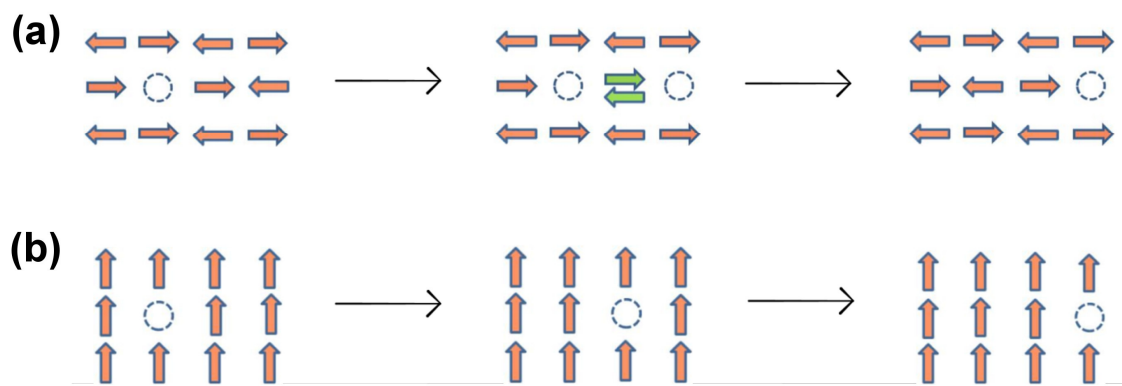

Supporting Fig. 5: Correlated hopping process in CeIn$_3$. **a** shows a schematic of the effective next nearest neighbour hopping of holes ($t'$ in the Hubbard model) in an antiferromagnet involving the exchange of two fermions and the creation of a virtual singlet. **b** shows a schematic of the direct hopping of holes which is favoured over the creation of virtual singlets once the $f$-electron spins are polarised in strong magnetic fields.